\documentclass{aa}
\usepackage{graphicx}
\usepackage{color}
\usepackage{txfonts}
\usepackage{longtable}
\usepackage{supertabular}
\usepackage{natbib}
\defcitealias{vanderBurg+14}{vdB14}
\begin{document}

\definecolor{green}{rgb}{0,0.5,0}
\definecolor{grey}{rgb}{0.4,0.5,0.7}
\newcommand{\ab}[1]{\textcolor{red}{\bf Andrea: #1}}
\newcommand{\lang}[1]{\textcolor{red}{\bf #1}}
\newcommand{\rev}[1]{\textcolor{magenta}{\bf #1}}
\newcommand{\msun}{M_{\odot}}
\newcommand{\ks}{\mathrm{km~s}^{-1}}
\newcommand{\rv}{r_{\Delta}}
\newcommand{\mv}{M_{\Delta}}
\newcommand{\cv}{c_{\Delta}}
\newcommand{\rtwo}{r_{200}}
\newcommand{\rtwoi}{r_{200,i}}
\newcommand{\vtwo}{v_{200}}
\newcommand{\rtwof}{r_{200}^{(0)}}
\newcommand{\rfive}{r_{500}}
\newcommand{\rtwofive}{r_{2500}}
\newcommand{\rume}{r_{\rm{200,U}}}
\newcommand{\mtwo}{M_{200}}
\newcommand{\ctwo}{c_{200}}
\newcommand{\rhor}{\rho(r)}
\newcommand{\br}{\beta(r)}
\newcommand{\rs}{r_{-2}}
\newcommand{\rh}{r_{\rm H}}
\newcommand{\rb}{r_{\rm B}}
\newcommand{\ri}{r_{\rm I}}
\newcommand{\rc}{r_{\rm c}}
\newcommand{\rn}{r_{\nu}}
\newcommand{\rr}{r_{\rho}}
\newcommand{\rg}{r_{\rm{g}}}
\newcommand{\rsg}{r_{\rm{-2,g}}}
\newcommand{\ra}{r_{\beta}}
\newcommand{\slos}{\sigma_{\rm{los}}}
\newcommand{\slosg}{\sigma_{\rm{los,g}}}
\newcommand{\sigv}{\sigma_v}
\newcommand{\qrr}{Q_r(r)}
\newcommand{\qr}{Q(r)}
\newcommand{\nr}{\nu(r)}
\newcommand{\vrf}{v_{{\rm rf}}}
\newcommand{\frs}{f_{{\rm sub}}}
\newcommand{\nm}{N_{{\rm m}}}
\newcommand{\ngr}{N_{{\rm g}}}
\newcommand{\cg}{c_{{\rm g}}}
\newcommand{\nmns}{N_{{\rm mns}}}
\newcommand{\ndyn}{N_{{\rm dyn}}}

\title{The concentration-mass relation of clusters of galaxies from
the OmegaWINGS survey}
%\subtitle{}
%\thanks{}

\author{
  A. Biviano\inst{\ref{ABi}} 
  \and
  A. Moretti\inst{\ref{AMo}}
  \and
  A. Paccagnella\inst{\ref{APa},\ref{AMo}}
  \and
  B. M. Poggianti\inst{\ref{AMo}}
  \and
  D. Bettoni\inst{\ref{AMo}}
  \and
  M. Gullieuszik\inst{\ref{AMo}}
  \and
  B. Vulcani\inst{\ref{BVu},\ref{AMo}}
  \and
  G. Fasano\inst{\ref{AMo}}
  \and
  M. D'Onofrio\inst{\ref{APa}}
  \and
  J. Fritz\inst{\ref{JFr}}
  \and
  A. Cava\inst{\ref{ACa}}
}

\offprints{A. Biviano, biviano@oats.inaf.it}

\institute{
  INAF-Osservatorio Astronomico di Trieste, via G. B. Tiepolo 11, 
I-34131, Trieste, Italy\label{ABi} 
\and
INAF-Astronomical Observatory of Padova,
vicolo dell'Osservatorio 5
35122 Padova, Italy\label{AMo}
\and
Department of Physics and Astronomy, University of Padova,
vicolo dell'Osservatorio 5, 35122 Padova, Italy\label{APa}
\and
School of Physics, The University of Melbourne, Swanston St \&
Tin Alley Parkville, VIC 3010, Australia\label{BVu}
\and
Instituto de Radioastronomia y Astrofisica, UNAM, Campus
Morelia, A.P. 3-72, C.P. 58089, Mexico\label{JFr}
\and
Department of Astronomy, University of Geneva, 51 Ch. des Maillettes, 1290 Versoix, Switzerland\label{ACa}
}
 
\date{August, 23$^{rd}$ 2017}

\abstract{The relation between a cosmological halo concentration and
  its mass ($cMr$) is a powerful tool to constrain
  cosmological models of halo formation and evolution.}{On the scale
  of galaxy clusters the $cMr$ has so far been determined mostly with
  X-ray and gravitational lensing data. The use of independent techniques
  is helpful in assessing possible systematics. Here we provide one of
  the few determinations of the $cMr$ by the dynamical analysis of the
  projected-phase-space distribution of cluster members.}{Based on the
  WINGS and OmegaWINGS data sets, we used the Jeans analysis with
the MAMPOSSt technique to determine masses and concentrations for 49 nearby clusters, each of which has $\gtrsim 60$ spectroscopic
  members within the virial region, after removal of
  substructures.}{Our $cMr$ is in statistical agreement with
  theoretical predictions based on $\Lambda$CDM cosmological
  simulations.  Our $cMr$ is different from most previous
  observational determinations because of its flatter slope and lower
  normalization. It is however in agreement with two recent $cMr$
  obtained using the lensing technique on the CLASH and LoCuSS cluster
  data sets.}{The dynamical study of the projected-phase-space of
  cluster members is an independent and valid technique to determine
  the $cMr$ of galaxy clusters. Our $cMr$ shows no tension with
  theoretical predictions from $\Lambda$CDM cosmological simulations
  for low-redshift, massive galaxy clusters. In the future we will extend
  our analysis to galaxy systems of lower mass and at higher redshifts.}


\titlerunning{$c-M$ relation of WINGS clusters}
\authorrunning{A. Biviano et al.}

\maketitle

\section{Introduction}
\label{s:intro}
The formation and evolution of dark-matter (DM) halos depend on the
cosmological model and are reflected in the halo internal
properties. The inner slope of a halo mass density profile $\rho(r)$ may be
sensitive to the DM properties \citep[e.g.,][]{YSWT00,CVAR08} and its
outer slope may carry information on the halo mass accretion rate
\citep{DK14}.  A full description of halo mass density profiles
requires a three-parameter model \citep{Navarro+04}, but a good
approximation is provided by the model of \citet[][NFW model
  hereafter]{NFW96},
\begin{equation}
\rho(r)=\frac{3 \, g(\cv) \, \mv}{4 \pi \, \rs^3 \, (r/\rs) \, (1+r/\rs)^2},
\end{equation}
with
\begin{equation}
g(\cv)=\frac{1}{\ln(1+\cv)-\cv/(1+\cv)},
\end{equation}
where $\rs$ is the radius at which $\rm{d} \ln \rho/\rm{d} \ln r = -2$,
$\cv \equiv \rv/\rs$, $\rv$ is the virial radius, related to 
the virial mass $\mv$ by
\begin{equation}
G \, \mv \equiv \Delta/2 \, H_z^2 \, \rv^3,
\label{e:m200}
\end{equation}
where $H_z$ is the Hubble constant at the halo redshift, $z$, and
$\Delta$ is the over-density with respect to critical.  The NFW model
is characterized by the two parameters, $\rs, \rv$, or equivalently,
$\cv, \mv$. These two parameters would specify the full evolution of a
halo in the spherical collapse model \citep{Bullock+01}.

Numerical simulations \citep[e.g.,][]{NFW96,Bullock+01} predict that
$\cv$ and $\mv$ are related by a relation ($cMr$ hereafter) that
evolves with $z$. The $cMr$ has a negative slope, that is, more
massive halos are less concentrated, and this is generally understood
as a direct consequence of the hierarchical accretion model for halo
formation and evolution. In fact, a halo concentration is related to
the ratio of the background density at the time of the first assembly
of its core mass and to the background density at the time the halo is
observed; in the hierarchical model, the first assembly epoch
occurs at higher $z$ , corresponding to a higher background density,
for lower mass halos \citep[e.g.,][]{Bullock+01,Dolag+04}. The $\cv$
dependence on $\mv$ is generally parametrized with a power law, i.e., $\cv
\propto \mv^a$ with a rather shallow slope, $a \approx -0.1$
\citep[e.g.,][]{NFW96,BHHV13}.

The distribution around the mean $cMr$ is lognormal with a
standard deviation that is related to the variance in the assembly histories of
DM halos \citep[e.g.,][]{Wechsler+02,ZJMB03,ZMJB03}. Less relaxed
halos are predicted to have smaller $\cv$ for given $\mv$ and a larger
scatter of the $cMr$ \citep[e.g.,][]{Jing00,Neto+07}.

At higher $z$, the $cMr$ is predicted to flatten and $\cv$ at given
$\mv$ is predicted to decrease
\citep[e.g.,][]{NFW96,Bullock+01,ZJMB03,Neto+07}.  While initial
studies based on cosmological simulations favored a strong dependence
of the $cMr$ on $z$, more recent works have predicted this dependence to be much shallower
with the normalization changing by $\sim 30$\% and the slope by $\sim
50$\% over the $z$ range 0--2 \citep[e.g.,][]{DeBoni+13,DM14}.  The
flattening of the $cMr$ with $z$ is attributed to the evolution of the
nonlinear mass scale and the transition from fast to slow assembly
mode; there is little evolution in $\cv$ when the mass growth rate is
fast \citep[e.g.,][]{ZJMB03,BHHV13,Correa+15-III}. As a result, little
evolution of $cMr$ is expected at the massive end because the
assembly epoch of massive halos is very recent
\citep[e.g.,][]{Fedeli12}.

The $cMr$, in particular its normalization and evolution, depends on
the cosmological model. Since $\cv$ at given $\mv$ is related to the
epoch of first halo assembly, models that change the rate of structure
formations also change the $cMr$ and its evolution. In this respect, the
most important parameters are the Hubble and density parameters $h$
and $\Omega_m$, the dispersion of the mass fluctuation within spheres
of comoving radius 8 $h^{-1}$ Mpc, $\sigma_8$, and the dark energy
equation of state parameter $w$
\citep[e.g.,][]{KMMB03,Dolag+04,MDvdB08,Carlesi+12,DeBoni+13,KBHH13}.

Given the information contained in the $cMr$ it is not surprising that
a considerable effort has been devoted to determine it from
observations, in particular at group and cluster mass scales
\citep[see][and references therein]{BHHV13,GGS16}.
Most of the $cMr$ determinations have been obtained either from X-ray
\citep{PAP05,Vikhlinin+06,Buote+07,Ettori+10,AECS16,MAM16} or from
lensing measurements \citep{CN07,MSH08,Covone+14,Umetsu+14,Umetsu+16,Du+15,Merten+15,vanUitert+16}. 

Comparing these determinations to the results of numerical simulations
has however proven not to be straightforward. On the numerical side,
baryonic physics must be included in the simulations. However, this
has little effect on the $cMr$ at the cluster scale; $\cv$
  increases by $\sim 10$\% in hydrodynamical simulations compared to
  the DM-only simulations \citep[see, e.g.,][]{Duffy+10,DeBoni+13}. Observational
effects may be more important than numerical effects in affecting the
$cMr$. The observed $cMr$ can be artificially steepened by the error
covariance in the measurements of $\cv$ and $\mv$ \citep{Auger+13},
unless this covariance is properly accounted for \citep{MAM16}.
Forcing an NFW model when this is not an adequate fit to the cluster
shear profiles also tends to steepen the $cMr$ \citep{SFM16}. The
observational selection of dynamically relaxed systems may be
different from that in cosmological simulations and this can change
the normalization and scatter of the $cMr$ \citep{Correa+15-III}.
Selecting a sample of clusters for their high X-ray luminosity or for
their strong lensing signal results in a steeper observed $cMr$ with
a higher normalization than that of the general population
\citep{Rasia+13,Giocoli+14,Meneghetti+14}. \citet{Rasia+13} has also
found that the hydrostatic assumption in the determination of cluster
mass profiles from X-ray data introduces a bias in the estimation of
both $\cv$ and $\mv$.

Given the possible systematics that can affect the $cMr$
determination, it is important to consider several cluster samples as
well as different methodologies. Both $\mv$ and $\cv$ can be
determined from the projected phase-space distribution of galaxies in
clusters and/or groups, but only a few studies have so far adopted
this approach to determine, or at least constrain, the $cMr$
\citep{Lokas+06,RD06,WL10}. In this paper we use data from the
WIde-field Nearby Galaxy-cluster Survey \citep[WINGS;][]{Fasano+06} and
its extension, OmegaWINGS \citep{Gullieuszik+15,Moretti+17} to
determine the mass density profiles and the $cMr$ of 49 nearby
clusters entirely from the projected phase-space distributions of
their member galaxies via the \texttt{MAMPOSSt} technique
\citep{MBB13}. This technique solves the Jeans equation for
  dynamical equilibrium \citep{BT87} by finding the parameters of
  given models for the mass and velocity anisotropy profiles, which
  maximize the combined probability of observing the projected
  phase-space distribution of cluster galaxies.

The structure of this paper is the following.  In Sect.~\ref{s:sample}
we describe our data set (Sect.~\ref{ss:data}), the selection of
cluster member galaxies, and the identification and removal of
substructures (Sect.~\ref{ss:member}) based on a new algorithm that
we describe in Appendix~\ref{a:ds}.  In Sect.~\ref{s:mprof} we
determine the cluster mass profiles that we use to derive the
$cMr$ in Sect.~\ref{s:cmr}. In the same Sect.~\ref{s:cmr} we compare
our $cMr$ to theoretical and other observational estimates of the
$cMr$.  We discuss our results in Sect.~\ref{s:disc} and provide a
summary of our results and our conclusions in Sect.~\ref{s:conc}.  

Throughout this paper we adopt the following cosmological parameter
values: a Hubble constant $H_0=70\,{\rm km\,s}^{-1} {\rm Mpc}^{-1}$, a
present-day matter density $\Omega_{\mathrm{m}}=0.3$, and a curvature
parameter value $\Omega_{k}=0$.

\section{The sample}
\label{s:sample}
\subsection{The data set}
\label{ss:data}
The WIde-field Nearby Galaxy Cluster Survey (WINGS) is a
multiwavelength survey of 76 clusters of galaxies in the redshift
range $0.04<z<0.07$ \citep{Fasano+06,Moretti+14}, X-ray selected from
the ROSAT All Sky Survey data \citep{Ebeling+96}. The WINGS clusters have
been imaged in the $B, V$ bands \citep{Varela+09}, and a subset
  of these clusters have been followed up with WYFFOS/WHT and 2dF/AAT spectroscopic
observations \citep{Cava+09}.  The OmegaWINGS \citep{Gullieuszik+15} is an
extension of WINGS both in terms of imaging and spectroscopy. Forty-six WINGS
clusters have been imaged with OmegaCAM/VST in the $u, B,$ and $V$
bands over areas of $\sim 1$ deg$^2$ each.  Thirty-three of these clusters have
been followed up with extensive spectroscopy with AAOmega/AAT
\citep[][D'Onofrio et al., in prep.]{Moretti+17}.

Galaxy redshifts have been measured from the spectroscopic
observations using a semi-automatic method, which involves the
cross-correlation technique and the emission lines identification,
with a success rate $\approx 95$\% down to an apparent magnitude
limit $V=20$
\citep{Cava+09,Fritz+11,Moretti+17}.

The WINGS and OmegaWINGS data have been complemented with data from
the literature, taken from SDSS/DR7 (603), NOAO (5), SIMBAD
  (1965), and NED (18721). In particular, redshift information from
  the cited catalogs were added for galaxies belonging to the parent
  photometric catalog that has been used for the WINGS/OmegaWINGS
  spectroscopic follow up, to allow the completeness estimation.

The completeness of the redshift catalog has been estimated for each
cluster as a function of both galaxy $V$ magnitudes and their
distances from their cluster center, which is defined as the position of the
brightest cluster galaxy (BCG); see \citet{Cava+09,Moretti+17}
  for details.

\subsection{Selection of cluster members}
\label{ss:member}
To identify cluster members we proceeded as follows. We first rejected
as obvious line-of-sight interlopers those galaxies in the cluster
field with $c \mid z-z_c \mid > 6000 \, \ks$ where $c$ is the speed of
light and $z_c$ is the first-guess cluster mean redshift, taken from
\citet{Fasano+06}.  We then applied the kernel mean matching (KMM)
algorithm \citep{McLB88,ABZ94} to look for the presence of multiple
peaks in the remaining $z$ distribution.  The KMM algorithm fits a
user-specified number of Gaussian distributions to a data set and
returns the probability that the fit by many Gaussians is better than
the fit by a single Gaussian. We always considered the simplest case
of only two Gaussians. When KMM indicated that a two-Gaussian fit is
better than a single-Gaussian fit with a probability of $\geq 0.95$, we
selected the most populated of the two Gaussians as our fiducial
cluster sample.  More specifically, we rejected from the sample those
galaxies that have a higher probability of being part of the less
populated Gaussian than of being part of the more populated Gaussian.
%% We then iterated the procedure on the
%% $z$ distribution of the remaining galaxies, but in no cluster was a
%% further subsample selection required according to the KMM test.
By this procedure we therefore identified the main peak of the cluster
in the $z$ distribution \citep{Beers+91,Girardi+93}.

In the second part of the procedure we removed additional interlopers
identified either by the \texttt{Shifting Gapper}
method \citep{Fadda+96} or by the \texttt{Clean}
method \citep{MBB13}, or by both methods. Both methods
identify interlopers based on their location in projected phase-space
$R, \vrf$, where $R$ is the projected radial distance from the cluster
center, $\vrf \equiv c \, (z-\overline{z})/(1+\overline{z})$ is the
rest-frame velocity, and $\overline{z}$ is the average redshift of the
members that have been selected in the first part of the membership
procedure.  For the \texttt{Shifting Gapper} method we adopted the
following parameters: 600 kpc for the bin size, a minimum of 15
galaxies per bin, and 1000 $\ks$ for the significance of the gap in
velocity space \citep[the meaning of these parameters is described in
  detail in][]{Fadda+96}.

For the remaining cluster members we searched for possible substructures
using a new procedure (\texttt{DS+}) detailed in Appendix~\ref{a:ds},
which we developed from a modification of the procedure of
\citet{DS88}. Results from the application of this test to the
  OmegaWINGS data set have already been used in \citet{Paccagnella+17}.
Galaxies with a formal probability of $ \geq 0.995$ of belonging to a
subcluster are rejected from the sample of cluster members.

Based on the sample of cluster members, we computed the mean cluster
redshift $z_c$ and the line-of-sight velocity dispersion, $\slos$, in
the rest frame of each cluster, using the robust biweight scale
estimator \citep[][we also adopt this estimator in the rest of this
  paper]{BFG90}. We then provided an initial estimate of the virial
radius $\rtwoi$ from $\slos$ using a scaling relation derived for NFW
models with velocity anisotropy estimated in \cite{MBM10}.

In Table~\ref{t:clist} we list the cluster name in Col.~1; the RA and declination of the BCG (adopted as cluster center) in Cols.~2
and 3;
the number of galaxies with $z$ in the cluster field, $N_z$, in Col.~4; 
the number of cluster members before the removal of galaxies
in subclusters, $\nm$, in Col.~5; and the final number of cluster
members after the removal of galaxies in subclusters, $\nmns$, in Col.~6.  In
Col.~7 we then list the number of cluster members effectively used in
the dynamical analysis described in Sect.~\ref{s:mprof}, $\ndyn$,
i.e., those located between 0.05 Mpc and $\rtwoi$. In Col.~8 we list
the largest distance from the BCG among the cluster members outside
substructures, ($R_{max}$), and in Cols.~9 and 10 we list the mean
redshift $z_c$, and velocity dispersion $\slos$, of the cluster.
Errors on $\slos$ are computed according to Eq.~(16) in \citet{BFG90}.

We only list in Table~\ref{t:clist} those 49 clusters with $\ndyn \geq
57$, since it was shown by \citet{Biviano+06}, based on a study of
cluster-size halos from cosmological simulations, that $\sim 60$ is
the minimum number of members to achieve, on average, an unbiased
estimate of cluster mass. We also exclude the cluster
A3530 from our sample  because its sample of members cannot be cleanly defined because of
its proximity to the more massive cluster A3532 \citep{LSSH13}.

A comparison of our $\slos$ determinations with those listed in
\citet{Moretti+17} for the 30 clusters in common indicates that
the latter are on average $10 \pm 2$\% higher. We attribute this
difference to the more accurate membership determination performed
in the present analysis.

\begin{table*}
  \centering
\caption{Cluster sample}
\label{t:clist}
\resizebox*{!}{0.95\textheight}{% resize table and text
\begin{tabular}{lrrrrrrcrrccccc}
\hline 
\\[-0.2cm]
Id  & RA & Dec &  $N_z$ & $\nm$ & $\nmns$ & $\ndyn$ & $R_{max}$ & $z_c$ & $\slos$ & $n(R)$ & $\rg$ & $M(r)$ & $\rtwo$ & $\rs$ \\
    & [deg] & [deg]  &  &       &         &         & [Mpc]     &      & [$\ks$] & model & [Mpc] & model  & [Mpc]   & [Mpc] \\[0.13cm]
\hline
\\[-0.1cm]
       A85 &  10.36130 &  -9.30300 &   1050 &  372 & 291 & 226 & 3.74 &   0.05568 & $  859_{ -44}^{+  42}$ &   King & $  0.48_{-  0.06}^{+  0.03}$ &   NFW & $    2.02_{-  0.20}^{+  0.13}$ & $    1.71_{-  0.60}^{+  2.86} $ \\[0.13cm]
      A119 &  13.98960 &  -1.26390 &    966 &  395 & 290 & 261 & 3.14 &   0.04436 & $  952_{ -49}^{+  46}$ &   King & $  0.24_{-  0.02}^{+  0.03}$ &   Her & $    2.25_{-  0.10}^{+  0.20}$ & $    0.52_{-  0.16}^{+  0.70} $ \\[0.13cm]
      A151 &  17.10920 & -15.40920 &   1023 &  294 & 207 & 149 & 3.68 &   0.05327 & $  771_{ -39}^{+  37}$ &   pNFW & $  0.35_{-  0.07}^{+  0.09}$ &   Bur & $    1.67_{-  0.09}^{+  0.10}$ & $    0.37_{-  0.10}^{+  0.19} $ \\[0.13cm]
      A160 &  18.16380 &  15.50740 &    474 &  120 &  84 &  70 & 3.04 &   0.04317 & $  738_{ -55}^{+  59}$ &   pNFW & $  0.27_{-  0.09}^{+  0.14}$ &   Her & $    1.60_{-  0.17}^{+  0.29}$ & $    0.72_{-  0.27}^{+  1.44} $ \\[0.13cm]
      A168 &  18.78250 &   0.28530 &   1277 &  231 & 150 &  89 & 3.35 &   0.04518 & $  498_{ -28}^{+  29}$ &   King & $  0.66_{-  0.15}^{+  0.22}$ &   Bur & $    0.97_{-  0.14}^{+  0.11}$ & $    0.33_{-  0.19}^{+  0.48} $ \\[0.13cm]
      A193 &  21.18170 &   8.70060 &    376 &  155 & 131 & 114 & 3.32 &   0.04852 & $  758_{ -45}^{+  48}$ &   King & $  0.21_{-  0.03}^{+  0.04}$ &   Bur & $    1.58_{-  0.09}^{+  0.14}$ & $    0.28_{-  0.11}^{+  0.17} $ \\[0.13cm]
      A376 &  41.39420 &  36.90330 &    222 &  164 & 119 & 116 & 3.30 &   0.04752 & $  832_{ -52}^{+  56}$ &   King & $  0.29_{-  0.04}^{+  0.05}$ &   Bur & $    1.66_{-  0.08}^{+  0.16}$ & $    0.20_{-  0.10}^{+  0.20} $ \\[0.13cm]
      A500 &  69.70250 & -22.10020 &    580 &  236 & 194 & 123 & 4.58 &   0.06802 & $  660_{ -33}^{+  34}$ &   King & $  0.33_{-  0.04}^{+  0.05}$ &   Bur & $    1.80_{-  0.15}^{+  0.20}$ & $    0.79_{-  0.23}^{+  0.77} $ \\[0.13cm]
      A671 & 127.12790 &  30.43260 &    520 &  169 & 118 &  88 & 3.55 &   0.04939 & $  730_{ -46}^{+  49}$ &   pNFW & $  0.29_{-  0.08}^{+  0.10}$ &   Her & $    1.49_{-  0.08}^{+  0.16}$ & $    0.18_{-  0.07}^{+  0.18} $ \\[0.13cm]
      A754 & 137.12920 &  -9.63040 &    936 &  517 & 409 & 333 & 3.76 &   0.05445 & $  816_{ -42}^{+  39}$ &   pNFW & $  0.67_{-  0.08}^{+  0.09}$ &   Bur & $    1.66_{-  0.06}^{+  0.13}$ & $    0.36_{-  0.08}^{+  0.15} $ \\[0.13cm]
     A957x & 153.40670 &  -0.92510 &   1487 &  167 & 116 &  86 & 3.18 &   0.04496 & $  631_{ -40}^{+  43}$ &   King & $  0.19_{-  0.03}^{+  0.03}$ &   Bur & $    1.42_{-  0.09}^{+  0.20}$ & $    0.40_{-  0.16}^{+  0.29} $ \\[0.13cm]
      A970 & 154.39000 & -10.67640 &    495 &  219 & 150 & 116 & 4.00 &   0.05872 & $  749_{ -42}^{+  44}$ &   pNFW & $  0.28_{-  0.06}^{+  0.08}$ &   NFW & $    1.63_{-  0.07}^{+  0.19}$ & $    0.28_{-  0.10}^{+  0.33} $ \\[0.13cm]
     A1069 & 159.92630 &  -8.68770 &    597 &  152 & 107 &  66 & 4.06 &   0.06528 & $  542_{ -36}^{+  38}$ &   King & $  0.41_{-  0.08}^{+  0.10}$ &   Her & $    1.18_{-  0.16}^{+  0.17}$ & $    0.57_{-  0.26}^{+  1.43} $ \\[0.13cm]
    A1631a & 193.20630 & -15.40180 &   1223 &  506 & 338 & 199 & 3.21 &   0.04644 & $  715_{ -36}^{+  35}$ &   King & $  0.93_{-  0.11}^{+  0.16}$ &   Her & $    1.39_{-  0.23}^{+  0.19}$ & $    3.80_{-  1.43}^{+  6.50} $ \\[0.13cm]
     A1644 & 194.28370 & -17.39910 &    434 &  313 & 256 & 230 & 3.26 &   0.04691 & $  945_{ -48}^{+  46}$ &   pNFW & $  0.34_{-  0.06}^{+  0.07}$ &   Bur & $    1.89_{-  0.06}^{+  0.13}$ & $    0.29_{-  0.09}^{+  0.15} $ \\[0.13cm]
     A1795 & 207.21420 &  26.59270 &    670 &  245 & 191 & 127 & 4.36 &   0.06291 & $  731_{ -36}^{+  38}$ &   King & $  0.24_{-  0.04}^{+  0.04}$ &   Her & $    1.72_{-  0.09}^{+  0.16}$ & $    0.48_{-  0.17}^{+  0.45} $ \\[0.13cm]
     A1983 & 223.24290 &  16.70800 &    619 &  221 & 143 &  79 & 3.09 &   0.04517 & $  407_{ -23}^{+  25}$ &   pNFW & $  0.55_{-  0.19}^{+  0.31}$ &   Bur & $    0.95_{-  0.09}^{+  0.08}$ & $    0.27_{-  0.10}^{+  0.22} $ \\[0.13cm]
     A1991 & 223.62830 &  18.64310 &    616 &  180 & 118 &  57 & 4.06 &   0.05860 & $  570_{ -36}^{+  38}$ &   King & $  0.21_{-  0.07}^{+  0.07}$ &   NFW & $    1.33_{-  0.13}^{+  0.20}$ & $    0.32_{-  0.16}^{+  1.38} $ \\[0.13cm]
     A2107 & 234.90830 &  21.77830 &    491 &  190 & 139 &  74 & 2.92 &   0.04166 & $  519_{ -30}^{+  32}$ &   King & $  0.09_{-  0.04}^{+  0.03}$ &   Her & $    1.15_{-  0.09}^{+  0.20}$ & $    0.13_{-  0.05}^{+  0.14} $ \\[0.13cm]
     A2124 & 236.24130 &  36.10990 &    609 &  193 & 139 &  93 & 4.55 &   0.06692 & $  733_{ -43}^{+  45}$ &   pNFW & $  6.0_{-  2.5}^{+ 24.}$ &   NFW & $    1.09_{-  0.16}^{+  0.06}$ & $    0.25_{-  0.09}^{+  0.37} $ \\[0.13cm]
     A2382 & 327.96710 & -15.69560 &    792 &  370 & 226 & 200 & 4.40 &   0.06442 & $  807_{ -41}^{+  39}$ &   King & $  0.39_{-  0.04}^{+  0.04}$ &   Bur & $    1.73_{-  0.08}^{+  0.10}$ & $    0.57_{-  0.26}^{+  0.42} $ \\[0.13cm]
     A2399 & 329.29710 &  -7.82230 &   1233 &  329 & 215 & 162 & 3.91 &   0.05793 & $  662_{ -34}^{+  32}$ &   King & $  0.39_{-  0.05}^{+  0.07}$ &   Bur & $    1.55_{-  0.08}^{+  0.09}$ & $    0.42_{-  0.14}^{+  0.23} $ \\[0.13cm]
     A2415 & 331.34960 &  -5.59180 &    603 &  200 & 131 & 106 & 3.41 &   0.05791 & $  683_{ -41}^{+  43}$ &   King & $  0.78_{-  0.12}^{+  0.17}$ &   Bur & $    1.19_{-  0.18}^{+  0.27}$ & $    1.07_{-  0.45}^{+  0.80} $ \\[0.13cm]
     A2457 & 338.82040 &   1.48300 &    719 &  274 & 205 & 149 & 3.93 &   0.05889 & $  605_{ -31}^{+  29}$ &   King & $  0.48_{-  0.07}^{+  0.07}$ &   Bur & $    1.31_{-  0.11}^{+  0.16}$ & $    0.60_{-  0.22}^{+  0.44} $ \\[0.13cm]
     A2589 & 350.88630 &  16.77790 &    257 &  171 & 141 & 139 & 2.96 &   0.04217 & $ 1147_{ -66}^{+  70}$ &   King & $  0.42_{-  0.09}^{+  0.12}$ &   Her & $    2.75_{-  0.32}^{+  0.30}$ & $    1.73_{-  0.64}^{+  4.93} $ \\[0.13cm]
     A2593 & 350.98370 &  14.64730 &    610 &  273 & 198 & 117 & 2.97 &   0.04188 & $  523_{ -26}^{+  27}$ &   King & $  0.20_{-  0.04}^{+  0.04}$ &   Her & $    1.21_{-  0.06}^{+  0.11}$ & $    0.24_{-  0.10}^{+  0.29} $ \\[0.13cm]
     A2626 & 354.12210 &  21.14450 &    232 &   97 &  82 &  66 & 2.18 &   0.05509 & $  650_{ -49}^{+  53}$ &   King & $  0.14_{-  0.04}^{+  0.05}$ &   Bur & $    1.48_{-  0.09}^{+  0.17}$ & $    0.17_{-  0.07}^{+  0.17} $ \\[0.13cm]
     A2717 &   0.77710 & -35.92480 &    822 &  187 & 154 &  77 & 3.52 &   0.04989 & $  470_{ -26}^{+  27}$ &   King & $  0.23_{-  0.03}^{+  0.05}$ &   Bur & $    1.17_{-  0.08}^{+  0.11}$ & $    0.30_{-  0.11}^{+  0.24} $ \\[0.13cm]
     A2734 &   2.81750 & -28.84230 &   1034 &  267 & 216 & 135 & 4.34 &   0.06147 & $  588_{ -30}^{+  28}$ &   King & $  0.60_{-  0.07}^{+  0.08}$ &   Bur & $    1.38_{-  0.13}^{+  0.15}$ & $    0.65_{-  0.23}^{+  0.68} $ \\[0.13cm]
     A3128 &  52.54330 & -52.53700 &   1228 &  660 & 336 & 246 & 4.16 &   0.06033 & $  793_{ -40}^{+  38}$ &   King & $  0.76_{-  0.07}^{+  0.08}$ &   Bur & $    1.58_{-  0.16}^{+  0.15}$ & $    0.77_{-  0.34}^{+  0.51} $ \\[0.13cm]
     A3158 &  55.77040 & -53.65310 &    877 &  403 & 310 & 289 & 3.68 &   0.05947 & $  948_{ -48}^{+  46}$ &   pNFW & $  0.55_{-  0.08}^{+  0.10}$ &   Bur & $    1.94_{-  0.06}^{+  0.13}$ & $    0.39_{-  0.10}^{+  0.15} $ \\[0.13cm]
     A3266 &  67.77460 & -61.44360 &   1389 &  821 & 587 & 511 & 4.13 &   0.05915 & $ 1095_{ -56}^{+  53}$ &   King & $  0.41_{-  0.03}^{+  0.03}$ &   Bur & $    2.31_{-  0.06}^{+  0.11}$ & $    0.46_{-  0.10}^{+  0.13} $ \\[0.13cm]
     A3376 &  90.15290 & -40.03260 &    648 &  307 & 211 & 179 & 3.25 &   0.04652 & $  756_{ -39}^{+  37}$ &   pNFW & $  0.40_{-  0.07}^{+  0.15}$ &   NFW & $    1.65_{-  0.08}^{+  0.17}$ & $    0.49_{-  0.15}^{+  0.57} $ \\[0.13cm]
     A3395 &  96.88000 & -54.43740 &   1020 &  516 & 354 & 334 & 3.49 &   0.05103 & $ 1272_{ -65}^{+  62}$ &   King & $  0.56_{-  0.03}^{+  0.06}$ &   Bur & $    2.76_{-  0.11}^{+  0.27}$ & $    1.32_{-  0.28}^{+  0.44} $ \\[0.13cm]
    A3528a & 193.63040 & -29.37270 &   1304 &  435 & 330 & 241 & 3.76 &   0.05441 & $  891_{ -45}^{+  43}$ &   King & $  0.54_{-  0.03}^{+  0.04}$ &   NFW & $    1.88_{-  0.09}^{+  0.16}$ & $    0.50_{-  0.18}^{+  0.55} $ \\[0.13cm]
     A3532 & 194.32330 & -30.35400 &    660 &  393 & 267 & 147 & 3.89 &   0.05536 & $  662_{ -34}^{+  32}$ &   King & $  0.48_{-  0.04}^{+  0.14}$ &   NFW & $    1.55_{-  0.25}^{+  0.13}$ & $    1.54_{-  0.67}^{+  3.50} $ \\[0.13cm]
     A3556 & 201.00670 & -31.65900 &   1203 &  564 & 426 & 153 & 3.39 &   0.04796 & $  531_{ -27}^{+  26}$ &   pNFW & $  0.88_{-  0.22}^{+  0.37}$ &   Bur & $    1.10_{-  0.09}^{+  0.13}$ & $    0.29_{-  0.11}^{+  0.21} $ \\[0.13cm]
     A3558 & 201.97540 & -31.48480 &   1662 & 1126 & 691 & 522 & 3.58 &   0.04829 & $  910_{ -46}^{+  44}$ &   King & $  0.76_{-  0.05}^{+  0.05}$ &   Bur & $    1.95_{-  0.11}^{+  0.16}$ & $    1.03_{-  0.20}^{+  0.33} $ \\[0.13cm]
     A3560 & 202.95250 & -33.22480 &    937 &  343 & 256 & 227 & 3.27 &   0.04917 & $  799_{ -41}^{+  39}$ &   pNFW & $  0.40_{-  0.06}^{+  0.10}$ &   NFW & $    1.79_{-  0.08}^{+  0.15}$ & $    0.79_{-  0.22}^{+  0.87} $ \\[0.13cm]
     A3667 & 303.09170 & -56.81520 &   1313 &  705 & 474 & 441 & 3.71 &   0.05528 & $ 1031_{ -53}^{+  50}$ &   King & $  0.64_{-  0.04}^{+  0.04}$ &   NFW & $    2.22_{-  0.12}^{+  0.10}$ & $    1.12_{-  0.28}^{+  0.64} $ \\[0.13cm]
     A3716 & 312.86000 & -52.70700 &    773 &  447 & 291 & 232 & 3.14 &   0.04599 & $  753_{ -38}^{+  36}$ &   King & $  0.48_{-  0.03}^{+  0.07}$ &   NFW & $    1.72_{-  0.15}^{+  0.11}$ & $    1.49_{-  0.51}^{+  2.16} $ \\[0.13cm]
     A3809 & 326.72540 & -43.88870 &    985 &  255 & 204 & 120 & 4.02 &   0.06245 & $  499_{ -25}^{+  24}$ &   King & $  0.56_{-  0.08}^{+  0.12}$ &   Bur & $    1.04_{-  0.12}^{+  0.14}$ & $    0.45_{-  0.20}^{+  0.49} $ \\[0.13cm]
     A3880 & 336.95500 & -30.56390 &   1114 &  281 & 194 &  95 & 4.00 &   0.05794 & $  514_{ -25}^{+  27}$ &   pNFW & $  0.33_{-  0.10}^{+  0.13}$ &   NFW & $    1.20_{-  0.06}^{+  0.15}$ & $    0.13_{-  0.06}^{+  0.22} $ \\[0.13cm]
     A4059 & 359.22960 & -34.74770 &   1285 &  369 & 240 & 180 & 3.40 &   0.04877 & $  744_{ -38}^{+  36}$ &   pNFW & $  0.55_{-  0.10}^{+  0.14}$ &   Bur & $    1.58_{-  0.09}^{+  0.14}$ & $    0.42_{-  0.13}^{+  0.26} $ \\[0.13cm]
   IIZW108 & 318.38420 &   2.56420 &    598 &  185 & 161 & 116 & 3.35 &   0.04889 & $  575_{ -31}^{+  33}$ &   King & $  0.41_{-  0.05}^{+  0.07}$ &   Bur & $    1.30_{-  0.10}^{+  0.15}$ & $    0.53_{-  0.24}^{+  0.45} $ \\[0.13cm]
     MKW3s & 230.46170 &   7.70930 &    712 &  161 & 134 &  85 & 3.14 &   0.04470 & $  604_{ -36}^{+  38}$ &   King & $  0.22_{-  0.04}^{+  0.04}$ &   NFW & $    1.58_{-  0.18}^{+  0.22}$ & $    1.06_{-  0.43}^{+  3.44} $ \\[0.13cm]
     Z2844 & 150.64880 &  32.70560 &    536 &  104 &  86 &  58 & 3.53 &   0.05027 & $  425_{ -31}^{+  34}$ &   pNFW & $  0.31_{-  0.13}^{+  0.23}$ &   Bur & $    0.88_{-  0.11}^{+  0.15}$ & $    0.37_{-  0.18}^{+  0.49} $ \\[0.13cm]
     Z8338 & 272.70540 &  49.92160 &    140 &   94 &  94 &  83 & 1.74 &   0.04953 & $  658_{ -46}^{+  50}$ &   pNFW & $  0.31_{-  0.11}^{+  0.16}$ &   Bur & $    1.35_{-  0.11}^{+  0.15}$ & $    0.36_{-  0.16}^{+  0.44} $ \\[0.13cm]
     Z8852 & 347.53170 &   7.58990 &    125 &   91 &  79 &  77 & 2.32 &   0.04077 & $  786_{ -60}^{+  65}$ &   King & $  0.23_{-  0.05}^{+  0.05}$ &   Bur & $    1.63_{-  0.10}^{+  0.29}$ & $    0.36_{-  0.13}^{+  0.32} $ \\[0.13cm]
\hline
\end{tabular}
}% end resize table and text
\tablefoot{  68\% upper and lower
uncertainties are listed for the velocity dispersion and for the
best-fit values of the $\rg, \rtwo,$ and $\rs$ parameters.}
\end{table*}

\section{Mass profiles}
\label{s:mprof}
We used \texttt{MAMPOSSt} in the so-called Split mode \citep[see
  Sect. 3.4 in][]{MBB13} to determine the mass profiles of the 49
selected WINGS clusters of Table~\ref{t:clist}. In the Split mode, the
number density profile of cluster galaxies, $n(R)$, is fit outside
\texttt{MAMPOSSt}. We used a weighted maximum-likelihood procedure to
fit the cluster $n(R)$ with two models: (1) a projected NFW model
\citep[pNFW hereafter; see][]{Bartelmann96}, and (2) a King model,
$n(R) \propto [1+(R/r_g)^2]^{-1}$ \citep{King62}. Both models are
characterized by two parameters: a scale radius that we call $\rg$,
where the `g' is for galaxies, and a normalization. However, the
$n(R)$ normalization is not a free parameter in the maximum-likelihood
procedure, since it is set by the requirement that the integrated
surface density over the cluster area equals the number of observed
members. In addition, the $n(R)$ normalization cancels out in the
dynamical analysis, so we do not consider this normalization here.

In fitting $n(R)$ to the radial distribution of cluster members we
weighed the galaxies by the inverse of the product of their radial and
luminosity incompleteness \citep{Moretti+17}. We only considered the
region between 0.05 Mpc and $\rtwoi$ for the fitting, for
  consistency with the radial limits chosen for the dynamical analysis
  (see below).

For each cluster we list in Col.~11 of Table~\ref{t:clist}, the model,
pNFW or King, which provides the best fit to the cluster $n(R)$, and in
Col.~12 the best-fit value of $\rg$ and its $1\sigma$
uncertainties. The King model provides a better fit to $n(R)$ than the
pNFW model in 33 of the 49 clusters considered. For the
  cluster A2124 the pNFW fit is preferred over the King fit, albeit
  with a very large, and essentially unconstrained, scale radius. This
  indicates that this cluster $n(R)$ is effectively a simple
  power law.

The preference of the King model over the pNFW model for $n(R)$
  confirms the results
  obtained by \citet{AMKB98} on a sample of 77 clusters from the
  ESO Nearby Abell Cluster Survey
\citep[ENACS,][]{Katgert+98}. \citet{LMS04} found instead that a pNFW model is
preferable over a cored model. The different findings can be explained
by the fact that cluster members in the \citet{LMS04} sample are
bright and $K$-band selected, and therefore contain a lower fraction
of blue, star-forming galaxies, which tend to avoid the central cluster
regions \citep[e.g.,][]{WGJ93,Biviano+97}. However, on an individual
cluster basis, the difference between the pNFW and the King fits is
generally not statistically significant. The pNFW (respectively King) model fit is only rejected in 9 (respectively 8) clusters with a
probability $>0.95$, according to a $\chi^2$ test. 

The best-fit model for $n(R)$ is deprojected with the Abel integral
\citep{BT87} to the 3D galaxy number density profile $\nu(r)$,
assuming spherical symmetry. Having fit the spatial distribution of
cluster members, we then used \texttt{MAMPOSSt} to fit their velocity
distribution. We only considered cluster members in the radial
  range 0.05 Mpc to $\rtwoi$. The lower radial limit is set to avoid
  the very central cluster region, dominated by the baryonic mass of
  the BCG \citep[e.g.,][]{BS06}, which is not included in our mass
  models (see Eqs. \ref{e:nfw}, \ref{e:her}, \ref{e:bur} below). The
  upper radial limit is set to avoid the cluster external regions,  which are less likely to have already attained dynamical
  equilibrium.

For each cluster galaxy we determined the probability of
observing its line-of-sight velocity in the cluster rest frame, at its
observed distance from the cluster center, given models for the mass
profile, $M(r)$, and the velocity anisotropy profile,
\begin{equation}
\beta(r) \equiv
1-(\sigma_{\theta}^2+\sigma_{\phi}^2)/(2 \, \sigma_r^2),
\end{equation}
where $\sigma_r$, $\sigma_{\theta}$, and $\sigma_{\phi}$ are the
radial, and the two tangential components, respectively, of the
velocity dispersion, and where we assumed
$\sigma_{\theta}=\sigma_{\phi}$. The best-fit parameters of the models
are those that maximize the product of all the cluster galaxy
probabilities. We used the \texttt{NEWUOA} Fortran code by
\citet{Powell06} to find the maximum likelihood, and then explored a
grid of parameter values around this maximum to set confidence limits.

We considered the following three models for $M(r)$:
\begin{enumerate}
\item The NFW model \citep{NFW97},
\begin{equation}
M(r)=\mtwo \frac{\ln(1+r/\rs)-r/\rs \, (1+r/\rs)^{-1}}{\ln(1+\ctwo)-c/(1+\ctwo)}
\ ,
\label{e:nfw}
\end{equation}
where $\ctwo \equiv \rtwo/\rs$, and $\rs$ is the radius where the logarithmic
derivative of the mass density profile equals $-2$.
\item The model of \citet{Hernquist90},
\begin{equation}
M(r)=\frac{\mtwo \, (\rh+\rtwo)^2}{\rtwo^2} \frac{r^2}{(r+\rh)^2}
\ ,
\label{e:her}
\end{equation}
Her model hereafter, where $\rh=2 \, \rs$.
\item The model of \citet{Burkert95},
\begin{eqnarray}
M(r) = \mtwo \, \{ \ln [1+(r/\rb)^2] + 2 \ln (1+r/\rb) \nonumber \\
- 2 \arctan (r/\rb) \} \times \{\ln [1+(\rtwo/\rb)^2] \nonumber \\
+ 2 \ln (1+\rtwo/\rb) - 2 \arctan (\rtwo/\rb) \}^{-1} \ ,
\label{e:bur}
\end{eqnarray} 
Bur model hereafter, where $\rb \simeq 2/3 \, \rs$. 
\end{enumerate}
The Bur and Her models differ from the NFW model because they are characterized
by a central core, and by a steeper asymptotic slope, respectively.
The virial and scale radii, $\rtwo$ and $\rs$, respectively,
are the two parameters characterizing these
models.

We considered three models for $\br$:
\begin{enumerate}
\item The C model, $\br=C$, in which the velocity anisotropy is
constant at all radii.
\item The OM model, $\br=r^2 \, (r^2+r_{\beta}^2)^{-1}$, which is characterized
  by the scale radius $r_{\beta}$ beyond which the anisotropy becomes
  increasingly more radial \citep{Osipkov79,Merritt85}.
\item The T model, $\br=\beta_{\infty} \, r \, (r+\rs)^{-1}$, which  is derived from a model introduced by \citet{Tiret+07}, and has been
  shown to fit the $\br$ of cluster-sized halos extracted from
  numerical simulations \citep{MBM10,MBB13}. Like the OM model, it is
  characterized by an increasingly radial velocity anisotropy with
  radius, but with a different functional form.
\end{enumerate}
All three $\br$ models contribute only one additional free parameter
to the \texttt{MAMPOSSt} analysis, i.e., $C, r_{\beta},$ or
$\beta_{\infty}$, since $\rs$ in the T model is the same parameter of
$M(r)$.

In Table~\ref{t:clist} we list the best-fit $M(r)$ model (Col.~13) and
its best-fit parameters $\rtwo$ and $\rs$ (Cols.~14 and 15), with 68\%
confidence limits obtained by marginalizing each parameter with
respect to the other two.  In general, constraints on the $\br$ model
are very poor and we prefer to omit these constraints here because they are
not particularly relevant in the context of the $cMr$.

\begin{figure}
\begin{center}
%\begin{minipage}{0.5\textwidth}
\resizebox{\hsize}{!}{\includegraphics{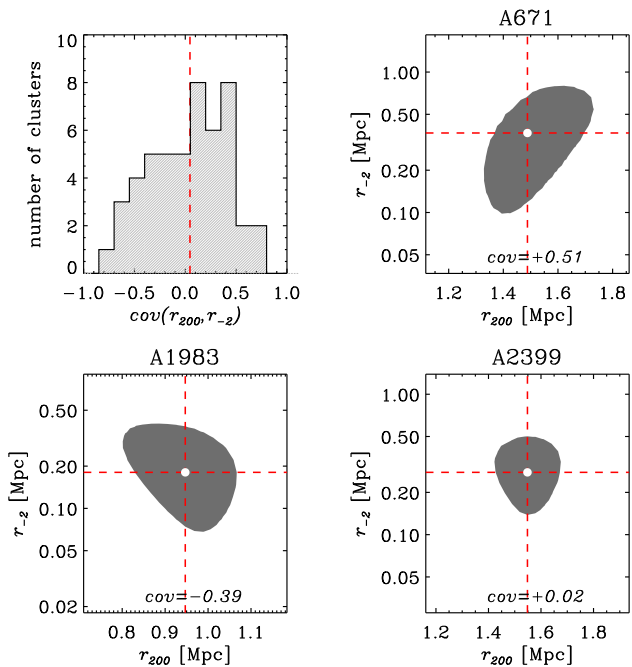}}
%\end{minipage}
\end{center}
\caption{Top left panel: distribution of error covariance
    between the $\rtwo$ and $\rs$ dynamical parameters. The mean of
    the distribution is indicated by a dashed vertical (red) line.
    Other panels: 68\% confidence contours and best-fit values of the
  $\rs$ (y-axis) vs. $\rtwo$ (x-axis) parameters of the
  \texttt{MAMPOSSt} analysis, in units of Mpc, for 3 of the
  49 clusters analyzed.  The best-fit solution is indicated by the
  dashed red lines and the white dot. The error
    covariance of the two parameters is listed at the bottom center of
    each panel.}  
\label{f:cov}
\end{figure}

In Fig.~\ref{f:cov} (top left panel) we show the
  distribution of the error covariance of the $\rtwo$ and $\rs$
  dynamical parameters for our 49 clusters. To estimate the error
  covariance for each cluster we determine the likelihoods $L_j$ of a
  grid of $\{\rtwo,\rs\}$ values around the best-fit solution. We then
  evaluate
\begin{equation}
  cov(\rtwo,\rs)= 
  \frac{\sum_{j=1}^n L_j \, (x_j-x_0) \times
    (y_j-y_0)}{[(\sum_{j=1}^n L_j \, (x_j-x_0)^2) \times ( \sum_{j=1}^n L_j \, (y_j-y_0)^2 ) ]^{1/2}} \ ,
    \label{e:cov}
\end{equation}
where $x_j=r_{200,j}$, and $y_j=r_{-2,j}$ are the $j=1, \ldots, n$ grid
  values, and $[x_0, y_0]$ is the best-fit \texttt{MAMPOSSt}
  solution for the two parameters. In the other three panels of Fig.~\ref{f:cov} we show
the 68\% confidence contours
in the $\rs$ versus $\rtwo$ plane for three 
clusters with values of $cov(\rtwo,\rs)$ representative of
  the full distribution.
  The error covariance distribution has a mean
  value that is consistent with zero, $<cov(\rtwo,\rs)>=0.05 \pm 0.06$.
In some clusters there is significant covariance of the
errors in the two parameters, but this is not generally the case,
and there is an almost equal fraction of clusters with
  positive and negative values of $cov(\rtwo,\rs)$.
In
this respect, the dynamical analysis based on cluster kinematics
differs from those based on X-ray and lensing, where the error
covariance of the $\ctwo$ and $\mtwo$ parameters tends to bias the
observed relation toward steeper slopes \citep{Auger+13,SGEM15}.

\begin{figure}
\begin{center}
%\begin{minipage}{0.5\textwidth}
\resizebox{\hsize}{!}{\includegraphics{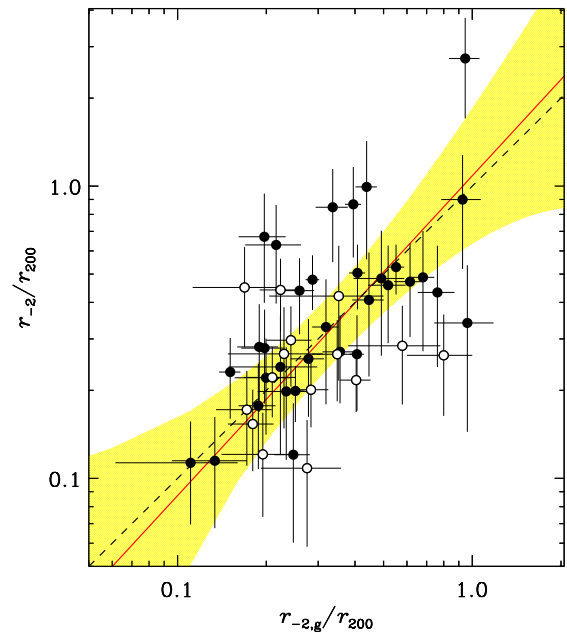}}
%\end{minipage}
\end{center}
\caption{Scale radii of the total cluster mass density profiles, in
  units of the virial radii ($\rs/\rtwo$ on the y-axis) vs. scale
  radii of the galaxy number density profiles, in the same units
  ($\rsg/\rtwo$ on the x-axis).  The error bars are 68\% confidence
  levels, but do not include the uncertainties on $\rtwo$, given that
  the same quantity is used to normalize both variables on the two
  axes. The dashed line indicates the identity relation $\rs \equiv
  \rsg$.  The solid (red) line indicates the best-fit log-log
  orthogonal relation between the two quantities, and the (yellow)
  shaded region indicates its 68\% confidence region, obtained via the
  \texttt{IDL} code \texttt{confidence\_band}.  The best fit was
  obtained via the fitting procedure of \citet{WBC10}. Open
  (respectively filled) dots indicate clusters whose $n(R)$ has been
  fit with a pNFW (respectively King) model. A2124 is an
    outlier on this relation and its $[\rsg/\rtwo,\rs/\rtwo]$ point
    lies off scale at $[5.5,0.23]$.}  
\label{f:r2r2g}
\end{figure}

We compare the scale radii of the galaxy and the total mass
distributions in Fig.~\ref{f:r2r2g}. In this figure we show the
best-fit values of $\rs/\rtwo$ versus $\rsg/\rtwo$ for our 49 clusters,
where $\rsg$ is the radius at which the logarithmic derivative of the
3D galaxy number density profile equals $-2$. For the pNFW model,
$\rsg=\rg$, while for the King model, $\rsg=\sqrt{2} \rg$. The inverse
of these quantities are the concentrations of the total mass and
galaxy distributions, $\ctwo$ and $\cg$, respectively. There is
a very significant correlation between $\rs/\rtwo$ and $\rsg/\rtwo$
\citep[Spearman rank correlation coefficient 0.51, corresponding to a
  probability of $2 \times 10^{-4}$; see e.g.,][]{Press+92}. A similar
correlation has been noted before by \citet[][see their
  Fig.~15]{vanUitert+16} on a large sample of cluster $\ctwo$ values
determined by a weak lensing stacking analysis.

We adopted the fitting procedure of \citet[][see their
  Eqs.~(3,4)]{WBC10}, which is based on the
minimization of $\chi^2$ defined by
\begin{equation}
\chi^2=\sum_{i=1}^N [y_i-(a x_i + b)]^2/\sigma_i^2
\label{e:chi2}
,\end{equation}
where the sum is over the $N=49$ clusters, $x$ and $y$ are the
logarithms of the observational data, $\sigma^2=\sigma_y^2+a^2
\sigma_x^2$, $a$ and $b$ are the intercept and slope of the fitted
relation. Following \citet{WBC10}, if $\chi^2/{\rm dof}>1$, where dof
is the number of degrees of freedom, 47 in our case (49 data - 2 free
parameters), an additional extra scatter term, $\sigma_{\rm{int}}$,
can be added in quadrature to $\sigma$, to lower the value of
$\chi^2$/dof to unity. Such an extra scatter term represents the
intrinsic scatter of the relation.

Since both x- and y-axis variables are affected by errors, and none
  of the two is dominating the observed scatter, we
  used the orthogonal distance regression \citep[see Sect. 4.2 in][]{FB92}.
  We find
\begin{equation}
  \log (\rs/\rtwo) = (0.0 \pm 0.1) + (1.1 \pm 0.2) \times \log (\rsg/\rtwo)
\label{e:r2r2g}
\end{equation}
with an intrinsic scatter around the relation of 0.23, which accounts
for 90\% of the total scatter. The mean ratio of the $\rs/\rsg$ values
for the 49 clusters is $0.92 \pm 0.08$, obtained using the robust
biweight estimator of central location; we adopted this estimator in
the rest of this paper as well \citep[][]{BFG90}. Our analysis
therefore indicates that, on average, galaxies are spatially
distributed like the total mass, in agreement with the result of
\citet[][see their Fig.~9]{BG03} taking into account that our analysis
is restricted to the virial region. \citet{vanUitert+16} found instead
$\cg > \ctwo$, on average, but they only considered red sequence
galaxies and these are known to be more centrally concentrated in
clusters than the whole cluster population because of the well-known
morphology-density relation \citep{Dressler80}.

\section{Concentration-mass relation}
\label{s:cmr}
\begin{figure}
\begin{center}
%\begin{minipage}{0.5\textwidth}
\resizebox{\hsize}{!}{\includegraphics{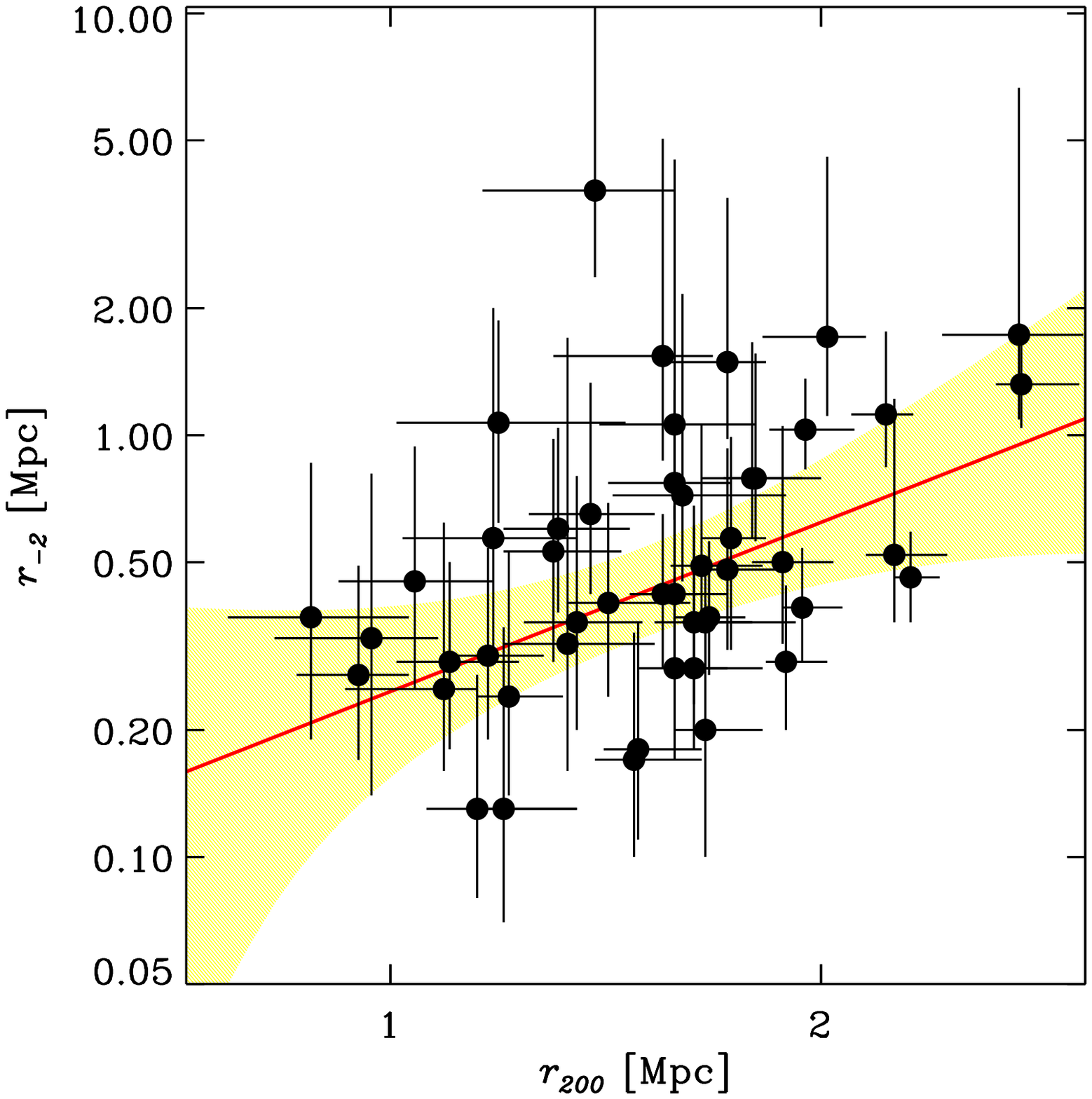}}
%\end{minipage}
\end{center}
\caption{Scale-radii $\rs$ vs. virial radii $\rtwo$, as obtained from
  the \texttt{MAMPOSSt} dynamical analysis, for the 49 clusters in our
  sample (Eq.~(\ref{e:rvrs})). Error bars indicate marginalized 68\%
  confidence levels on the measured values.  The solid (red) line
  indicates the best-fit log-log orthogonal relation between the two
  quantities, and the (yellow) shaded region its 68\% confidence
  region, obtained using the \texttt{IDL} code
  \texttt{confidence\_band}.  The best fit was obtained using the
  fitting procedure of \citet{WBC10}.}
\label{f:rvrs}
\end{figure}

In Fig.~\ref{f:rvrs} we show $\rs$ versus $\rtwo$ for our sample of 49
clusters.
There is a significant correlation between the two quantities
(Spearman rank correlation
coefficient 0.45, corresponding to a probability 0.001). Using the
fitting procedure of \citet{WBC10} and taking the
orthogonal relation, we find
\begin{equation}
\log \rs = (-0.61 \pm 0.08) + (1.3 \pm 0.3) \times \log \rtwo,
\label{e:rvrs}
\end{equation}
where both $\rtwo$ and $\rs$ are in Mpc.  From our data we measured
$\chi^2/{\rm dof}<1$ without adding the extra scatter (see the first two lines
of Table~\ref{t:chi2}), which
therefore remains undetermined.  In other words, our observational
uncertainties are too large to allow us to measure
$\sigma_{\rm{int}}$.

We used the relation of Eq.~(\ref{e:rvrs}) to derive the $cMr$,
\begin{equation}
\log \ctwo = (2.2 \pm 1.4) - (0.11 \pm 0.10) \times \log \mtwo,
\label{e:cmrder}
\end{equation}
where $\mtwo$ is in $\msun$ units. Alternatively, we derived the $cMr$
by direct fitting of
$\ctwo$ versus $\mtwo$ for our 49 clusters sample, using the
procedure of \citet{WBC10}, and we find the orthogonal relation
\begin{equation}
\log \ctwo = (1.0 \pm 1.4) - (0.03 \pm 0.09) \times \log \mtwo,
\label{e:cmr}
\end{equation}
which is fully consistent with the relation of Eq.~(\ref{e:cmrder}).
We did not apply any correction for error covariance
\citep{MAM16} because there is no systematic error covariance for the
$\ctwo$ and $\mtwo$ parameters in our cluster sample (see
Sect.~\ref{s:mprof} and Fig.~\ref{f:cov}).  The orthogonal
scatter in the relation of Eq.~(\ref{e:cmr}) is 0.22, and also in this
case it is dominated by observational uncertainties, since
$\chi^2/{\rm dof}<1$ (see Table~\ref{t:chi2}) without need for
including the extra intrinsic scatter term $\sigma_{\rm{int}}$.

\begin{figure}
\begin{center}
%\begin{minipage}{0.5\textwidth}
\resizebox{\hsize}{!}{\includegraphics{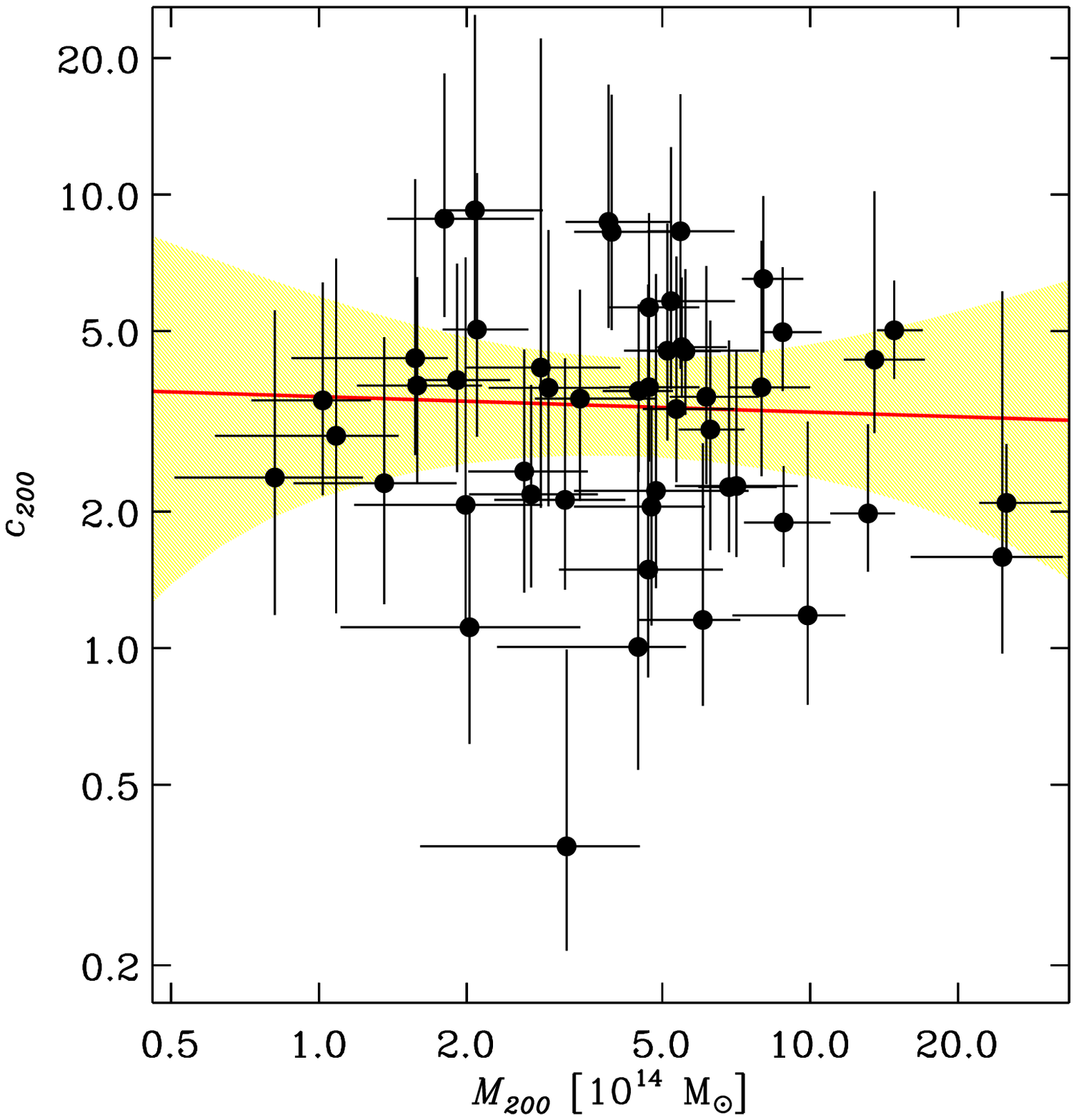}}
%\end{minipage}
\end{center}
\caption{Concentration $\ctwo$ vs. mass $\mtwo$, as obtained from the
  \texttt{MAMPOSSt} dynamical analysis, for the 49 clusters in our
  sample (Eq.~(\ref{e:cmr})). Error bars indicate 68\% confidence levels
  on the measured values.  The solid (red) line indicates the
  best-fit log-log orthogonal relation between the two quantities, and
  the (yellow) shaded region its 68\% confidence region, obtained using the
  \texttt{IDL} code \texttt{confidence\_band}.  The best fit
  was obtained using the fitting procedure of \citet{WBC10}.}
\label{f:cmr}
\end{figure}

\begin{figure}
\begin{center}
%\begin{minipage}{0.5\textwidth}
\resizebox{\hsize}{!}{\includegraphics{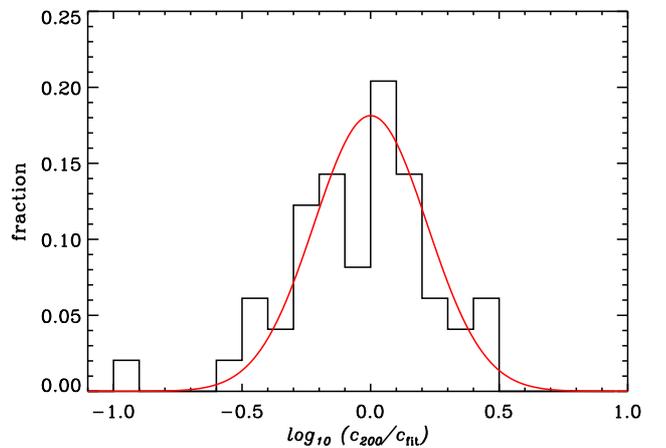}}
%\end{minipage}
\end{center}
\caption{Distribution of $\ctwo/c_{\rm{fit}}$ for the 49 clusters in
  our sample, where $c_{\rm{fit}}$ is the best-fit concentration at
  given $\mtwo$ from Eq.~(\ref{e:cmr}). The red curve is a lognormal
  curve centered at $\log \ctwo/c_{\rm{fit}}=0$, with a scatter of
  0.22, as obtained from the fitting procedure of \citet{WBC10}.}
\label{f:clogn}
\end{figure}

There is no significant correlation between $\ctwo$ and $\mtwo$
(Spearman rank correlation coefficient $-0.09$, corresponding to
  a probability 0.55), as expected given the flatness of the $cMr$
and the relatively large observational uncertainties. The $cMr$ of
Eq.~(\ref{e:cmr}) is shown in Fig.~\ref{f:cmr}, along with the
$\{\ctwo,\mtwo\}$ data points. In Fig.~\ref{f:clogn} we show the
distribution of $\ctwo/c_{\rm{fit}}$, where $c_{\rm{fit}}$ is the
best-fit concentration at given $\mtwo$ from
Eq.~(\ref{e:cmr}). Figure~\ref{f:clogn} shows that the distribution is
well fit by a lognormal curve, as expected theoretically
  \citep[see, e.g.,][]{Jing00,Bullock+01,Dolag+04} with a dispersion
of 0.22; this value is obtained from the $\ctwo$ versus $\mtwo$ fitting
procedure.

Different model choices for $n(R)$ and $M(r)$ (see
  Sect.~\ref{s:mprof}) have little effect on the $cMr$. We searched for
  systematic deviations from the $cMr$ of Eq.~\ref{e:cmr}, by
  evaluating the $\chi^2$ as in Eq.~(\ref{e:chi2}) separately for
  different subsets of clusters. Specifically we considered the five
  subsets of clusters selected according to their best-fit $n(R)$
  (King or pNFW) or $M(r)$ models (Bur, Her, or NFW). The derived
  $\chi^2$ values correspond to probabilities that imply no
  significant deviation from the $cMr$ of Eq.~\ref{e:cmr}.
  
Since many previous determination of the $cMr$ have been
  obtained by adopting the NFW $M(r)$ model, we redetermined
  the $cMr$ by forcing this model to all our clusters, and we
  find the orthogonal relation
\begin{equation}
  \log \ctwo = (1.2 \pm 2.0) - (0.05 \pm 0.14) \times \log \mtwo,
\label{e:cmrnfw}
\end{equation}
  fully consistent (albeit with larger scatter)
  with the $cMr$ of  Eq.~\ref{e:cmr}.
  
\begin{figure}
\begin{center}
%\begin{minipage}{0.5\textwidth}
\resizebox{\hsize}{!}{\includegraphics{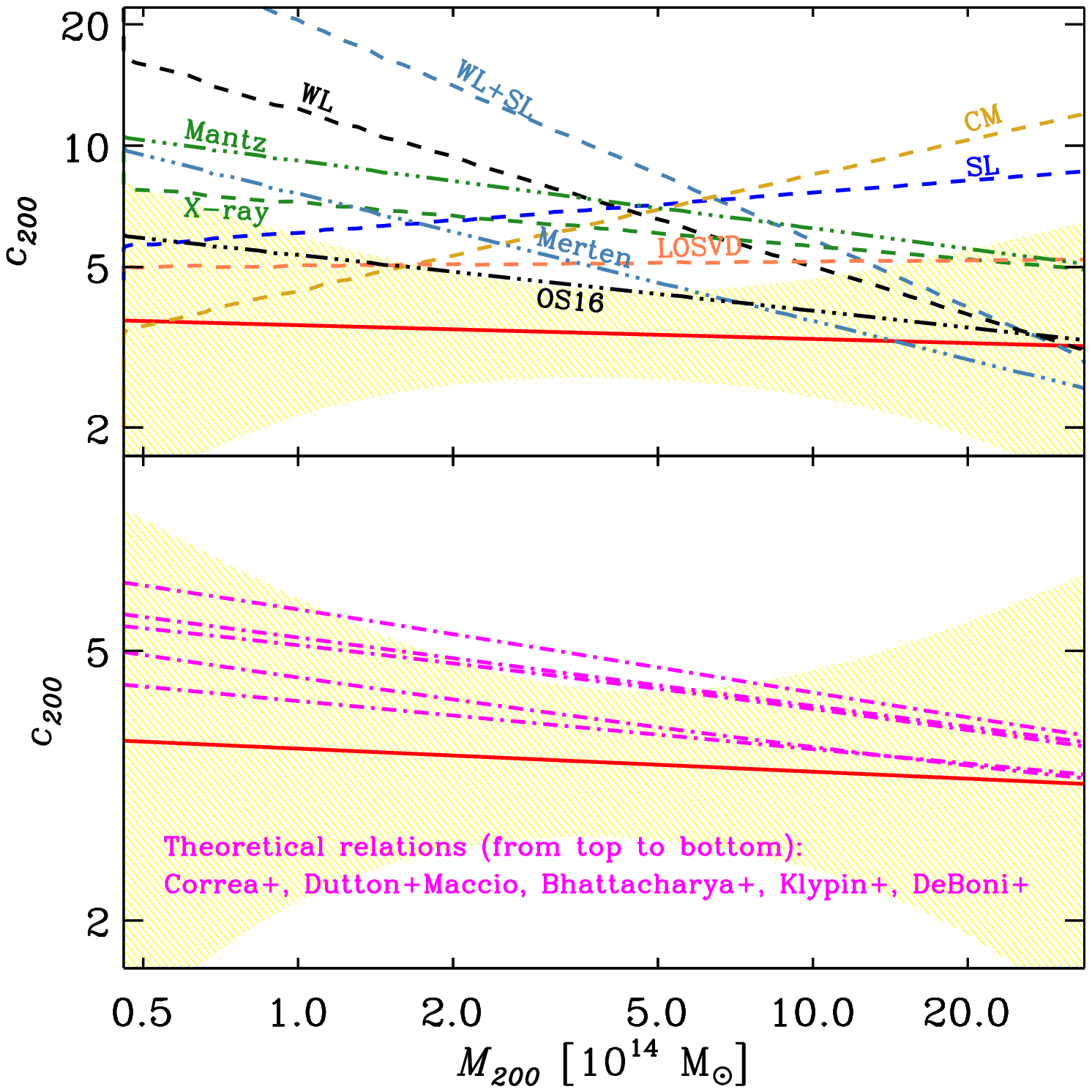}}
%\end{minipage}
\end{center}
\caption{Best-fit $cMr$ from Eq.~(\ref{e:cmr}) and Fig.~\ref{f:cmr}
  (red solid line; yellow shading indicates the 68\% confidence
  region), compared to other observational and theoretical $cMr$
  determinations. Upper panel: comparison with other
    observational determinations of the $cMr$:
  \citet[][triple-dot-dashed green line labeled 'Mantz']{MAM16},
  \citet[][tripled-dot-dashed steel blue line, labeled
    'Merten']{Merten+15}, \citet[][triple-dot-dashed black line,
    labeled 'OS16']{OS16}, and \citet[][dashed lines labeled as in
    Fig.~2 of their paper]{GGS16}. Bottom panel: comparison
    with theoretical determinations of the $cMr$;
  from top to bottom:
  \citet{Correa+15,DM14,BHHV13,Klypin+16,DeBoni+13}.}
\label{f:cfr}
\end{figure}

In Fig.~\ref{f:cfr} we compare the $cMr$ of Eq.~(\ref{e:cmr}) with
other observational (top panel)
and theoretical (bottom panel) estimates. The observational
estimates of \citet{GGS16} are based on several observational data and
techniques, namely, weak and strong lensing (labeled, respectively,
'WL' and 'SL' in the figure), hydrostatic equilibrium applied to X-ray
data (labeled 'X-ray' in the figure), the Jeans equation
\citep{Lokas02,LM03} or the Caustic technique \citep{DG97} applied to
the projected phase-space distribution of galaxies in the cluster
region (labeled LOSVD and CM in the figure, respectively). Other
displayed observational $cMr$ are from
\citet{Merten+15,MAM16,OS16} and based on weak+strong lensing,
X-ray, and weak lensing data, respectively.  The theoretical relations
shown in Fig.~\ref{f:cfr}
are from \citet{DeBoni+13,BHHV13,DM14,Correa+15,Klypin+16}. All $cMr$
are converted to an over-density of 200, when needed, assuming the NFW
mass profile, and evaluated at a redshift $z=0.052$, which is the
average redshift of our 49 clusters.

To estimate the level of agreement of the (theoretical or
observational) $cMr$ shown in Fig.~\ref{f:cfr}, with our own
$\{\ctwo,\mtwo\}$ data, we evaluate the goodness of the fits using
Eq.~(\ref{e:chi2}). In that equation, to take into account the
observational uncertainties of $cMr$ of other authors, we add in
quadrature an additional scatter term, $\sigma_{+}$, derived from the
uncertainties in the parameters of the different $cMr$, as given by
their authors. We neglect the uncertainties in the parameters of the
theoretical $cMr$; these are typically much smaller than the
uncertainties in the parameters of the observational $cMr$.  We
list the $\sigma_{+}$, the resulting $\chi^2$/dof values, and their
associated probabilities in Table~\ref{t:chi2}.  Our data favor a
low-$\ctwo$ normalization of the $cMr$, close to the theoretical $cMr$
of \citet{DeBoni+13}, but they are also in agreement with the other
theoretical relations shown in Fig.~\ref{f:cfr}. On the other hand,
many observational determinations of the $cMr$ disagree significantly
with our data, except those of \citet{Merten+15} and \citet{OS16}, and
the LOSVD and SL relations of \citet{GGS16}.

\begin{table}
\centering
\caption{Comparison with observational and theoretical $cMr$}
\label{t:chi2}
\begin{tabular}{lrrr}
\hline 
\\[-0.2cm]
Reference & $\sigma_{+}$ & $\chi^2$/dof & probability \\ [+0.1cm] 
\hline 
Eq.~(\ref{e:cmrder}) & -- & 0.88 & 0.70 \\
Eq.~(\ref{e:cmr}) & -- & 0.88 & 0.70 \\
\hline
\citet{GGS16} CM & 0.08 & 1.39 & $0.04$ \\
\citet{GGS16} LOSVD & 0.05 & 0.94 & 0.60 \\
\citet{GGS16} X-ray & 0.02 & 1.61 & $<0.01$ \\
\citet{GGS16} SL & 0.12 & 1.14 & $0.24$ \\
\citet{GGS16} WL & 0.03 & 1.94 & $<0.01$ \\
\citet{GGS16} WL+SL & 0.03 & 3.09 & $<0.01$ \\
\citet{MAM16} X-ray & 0.07 & 1.51 & $0.01$ \\
\citet{OS16} WL & 0.15 & 0.46 & $>0.99$ \\
\citet{Merten+15} WL+SL & 0.02 & 1.05 & 0.38 \\
\hline
\citet{Correa+15} & -- & 1.22 & 0.14 \\
\citet{DM14} & -- & 1.10 & 0.29 \\
\citet{BHHV13} & -- & 1.09 & 0.33 \\
\citet{Klypin+16} & -- & 0.93 & 0.61 \\
\citet{DeBoni+13} & -- & 0.91 & 0.65 \\
\hline
\end{tabular}
\tablefoot{Column 1 gives the reference to the $cMr$ used to compute
  the reduced-$\chi^2$ of Col. 2 with the observational $\{\ctwo,
  \mtwo\}$ data for our 49 clusters sample, using
  Eq.~(\ref{e:chi2}). Column 3 gives the probability that, in a
  $\chi^2$-distribution with 47 dof, a random variable is greater than
  or equal to the observed value listed in Col. 2; low values
  indicate that the $cMr$ is a poor representation of the
  observational data.}
\end{table}

\section{Discussion}
\label{s:disc}
Using the WINGS and OmegaWINGS data sets
\citep{Fasano+06,Gullieuszik+15,Poggianti+16,Moretti+17}, we
have derived $\ctwo$ and $\mtwo$ for 49 nearby clusters with $\gtrsim 60$
cluster members with positions and redshifts (see Tables~\ref{t:clist}
  and \ref{t:clist}). The determination of $\ctwo$ and $\mtwo$ was
obtained by application of the Jeans equation for dynamical
equilibrium of a spherical gravitating system via the
\texttt{MAMPOSSt} technique \citep{MBB13}. While not all clusters are
expected to be fully dynamically relaxed, \texttt{MAMPOSSt} has also been
shown to work reasonably well for partially relaxed clusters, in
particular if the analysis is restricted to the virial region, as is the case here.  To further reduce the effects of deviation from dynamical
relaxation, before running \texttt{MAMPOSSt}, we removed from our
sample those cluster members that were assigned a high probability of
belonging to cluster substructures based on our new \texttt{DS+}
procedure described in Appendix~\ref{a:ds}.

We find that mass and galaxies follow the same spatial distribution in
clusters, even if these were not forced to be the same in our
dynamical analysis.  More specifically, we find that the scale radii
(in units of $\rtwo$) of the galaxies and total matter distributions
are correlated (see Fig.~\ref{f:r2r2g}). This is the equivalent of saying
that the galaxies and total mass spatial concentrations ($\cg$ and
$\ctwo$) are correlated, which is a result originally obtained by
\citet{vanUitert+16}. Not only do we find $\cg$ and $\ctwo$ to be
correlated, but we also find that $\cg \approx \ctwo$ on average (see
Eq.~(\ref{e:r2r2g})). This finding confirms previous results
\citep{Carlberg+97-mprof,Mahdavi+99,Rines+01,BG03}. The scatter in the
observed $\rsg/\rtwo$ and $\rs/\rtwo$ relation is however substantial
and mostly intrinsic. We should therefore expect to find clusters
where galaxies and the mass have different distributions
\citep[e.g., MACS1206, or LCDCS0504,
  see][respectively]{Biviano+13,Guennou+14}, which is a warning against the
temptation to assume $\rs \equiv \rsg$ as a general rule in future
dynamical analyses.

We find a strong correlation between the \texttt{MAMPOSSt}-determined
values of $\rtwo$ and $\rs$ (see Fig.~\ref{f:rvrs}). Also in this
case, the relation is almost linear (see Eq.~(\ref{e:rvrs})). As a
consequence of the almost linear relation between $\rtwo$ and $\rs$,
$\ctwo \equiv \rtwo/\rs$ is almost independent of $\rtwo$, and
therefore -- via Eq.~(\ref{e:m200}) -- also almost independent of
$\mtwo$. In other words, the $cMr$ we obtain from the $\rtwo$
versus $\rs$ best-fitting relation, is nearly flat
(Eq.~(\ref{e:cmrder})).  This $cMr$ is fully consistent with the $cMr$
we obtain directly from fitting $\ctwo$ versus $\mtwo$ (Eq.~(\ref{e:cmr})
and Fig.~\ref{f:cmr}). We do not correct our $cMr$ for the
$\ctwo$ and $\mtwo$ error covariance \citep{Auger+13}, since this is
different for different clusters (see Sect.~\ref{s:mprof} and
  Fig.~\ref{f:cov}) and therefore unlikely to contribute a
systematic trend in the $\ctwo$ versus $\mtwo$ diagram of the full sample
of 49 clusters.

Our $cMr$ is in good agreement with theoretical predictions (see
Fig.~\ref{f:cfr} and Table~\ref{t:chi2}). In particular, the slope of
our $cMr$ range from -0.12 to 0.06, and the value of $c_{15}$ (i.e.,
$\ctwo$ at the mass scale $\mtwo=10^{15} \msun)$ ranges from 3.2 to
4.0.  We therefore favor a low-normalization $cMr$ such as that of
\citet{DeBoni+13}, but in fact all other theoretical $cMr$ considered
in this paper \citep{BHHV13,DM14,Correa+15,Klypin+16} are consistent
with our data. In other words, our data confirm theoretical
expectations about the $cMr$ of cosmological cluster-size halos, but
are unable to discriminate among different models. In this sense, our
results support the popular $\Lambda$CDM cosmological model
for the formation and evolution of cluster-size halos, but do not
constrain the parameters of this model to an accuracy comparable to
that allowed by other independent, observational constraints
such as those coming from cosmic microwave background
  observations \citep[see, e.g.][and references therein]{Correa+15}.

The uncertainties in our derived values of $\ctwo$ and $\mtwo$ are too
large to allow us to measure the intrinsic scatter in the $cMr$ we have
determined (see Fig.~\ref{f:cmr} and Eq.~(\ref{e:cmr})). We can only
provide an upper limit to the intrinsic scatter in the $cMr$, that is
the measured logarithmic scatter of 0.22 (0.51 in $\ln \ctwo$). Because this
value is an upper limit, it is consistent with the scatter measured
for the $cMr$ of halos from cosmological simulations \citep[$\sim
  0.2-0.3$; see, e.g.,][]{Jing00,Bullock+01,Dolag+04}. Our observed
distribution of $\log \ctwo/c_{\rm{fit}}$ is well represented by a
lognormal function, in line with theoretical predictions \citep[see,
  e.g.,][]{Jing00,Bullock+01,Dolag+04}.

Even if our $cMr$ is not of sufficient quality to allow
precision-cosmology determination of the cosmological parameters,
the good agreement between our observed $cMr$ and
$\Lambda$CDM theoretical parameters is very remarkable, given previous --
even recent -- claims of a significant tension between the
observational constraints and the numerical predictions of the $cMr$
\citep[e.g.,][]{SA07,CN07,Fedeli12,Du+15,vanUitert+16}. More
specifically, observed $cMr$ have frequently been found to be steeper
and of higher normalization than theoretical $cMr$. These discrepancies
are evident from Fig.~\ref{f:cfr}. In this figure, we compare our $cMr$ with
several $cMr$ obtained from numerical simulations
\citep{BHHV13,DeBoni+13,DM14,Correa+15,Klypin+16} and with
other observational determinations of the $cMr$ based on X-ray data
\citep{MAM16}, weak-lensing data \citep{OS16}, or a combination of
weak- and strong-lensing data \citep{Merten+15}, and on a variety of
data sets, including X-ray, weak- and strong-lensing, and
projected-phase-space galaxy distributions \citep{GGS16}.  Most of
these observational $cMr$ are steeper and/or have a higher
normalization than the theoretical predictions. Three observational
$cMr$ \citep[labeled CM, LOSVD, and SL in][and
  Fig.~\ref{f:cfr}]{GGS16} have an unexpected inverted slope
(i.e., $\ctwo$ increases with increasing $\mtwo$). Only the $cMr$ of
\citet{OS16}, based on weak-lensing data for 50 X-ray luminous
clusters from LoCuSS \citep{Okabe+13}, is in excellent agreement with
the theoretical predictions.

The comparison of our $cMr$ with other observational $cMr$ is
quantified in Table~\ref{t:chi2}.  Since our $cMr$ is consistent with
all the theoretical $cMr$, it is also inconsistent with those
observational $cMr$ that are most different from the theoretical  $cMr$.
Our $cMr$ is instead in good or reasonable agreement with the
observational $cMr$ of \citet{OS16} and \citet{Merten+15},
respectively; the latter is based on strong- and weak-lensing data for
clusters from the CLASH data set \citep{Postman+12}.  Our $cMr$
  is also consistent with the SL $cMr$ of \citet{GGS16}, but the
  agreement is due to the large uncertainties in the parameter values
  of this $cMr$.

Finally, it is particularly interesting to compare our $cMr$ with
the CM and LOSVD $cMr$ of Fig.~\ref{f:cfr} and \citet{GGS16}.
These two $cMr$ are both based on the analysis of the projected-phase
space distributions of cluster galaxies. More specifically, the
$\ctwo$ and $\mtwo$ values used for the CM $cMr$ are obtained by
application of the Caustic method \citep{DG97,Diaferio99}, while
those used for the LOSVD $cMr$ are obtained by the Jeans analysis
similar to our analysis here, but with different techniques
\citep{Lokas02,LM03} from \texttt{MAMPOSSt}. The CM and LOSVD
$cMr$ are based on 63, respectively 58, clusters \citep[see Table 2
  in][]{GGS16}, that is, a similar statistics as our data set. The
'CM' $cMr$ is statistically inconsistent with our $cMr$. This could be
related to the known bias of the Caustic method that leads to
over-estimating the cluster mass at small clustercentric distances, and
thereby $\ctwo$ \citep{Serra+11}. On the other hand, the 'LOSVD' $cMr$
is consistent with our $cMr$ (see Table~\ref{t:chi2}).

\section{Conclusions}
\label{s:conc}
We have derived the $cMr$ of 49 nearby clusters with data from the
WINGS and OmegaWINGS surveys
\citep{Fasano+06,Cava+09,Gullieuszik+15,Poggianti+16,Moretti+17}, by
applying the \texttt{MAMPOSSt} technique \citep{MBB13} to the
projected-phase-space distributions of cluster members. We used
particular care in defining cluster membership and developed a
partially new methodology (described in Appendix~\ref{a:ds}) to
identify and remove cluster substructures before performing the
dynamical analysis. While we did not force the mass distribution to be
identical to the distribution of cluster galaxies in our analysis,
this was indeed found to be the case, in an average sense, but
with significant scatter from cluster to cluster.

Our $cMr$ was found to be consistent with theoretical predictions from
$\Lambda$CDM cosmological simulations.  The $\ctwo/c_{{\rm fit}}$
distribution was found to follow a lognormal distribution, as
predicted theoretically. The quality of our $\ctwo$ and $\mtwo$
measurements is not sufficient to determine the intrinsic scatter of
the $cMr$, nor to constrain cosmological parameters to the level
required by the current precision cosmology. Nevertheless, we consider
it remarkable that our $cMr$ slope and normalization are so close to
the theoretical predictions, given that these have often been shown to
be in tension with observational $cMr$ determinations, in the recent
past. Only recently, more accurate $cMr$ determinations, based on
gravitational lensing, appear to have overcome the discrepancy with
the theoretical expectations and our $cMr$ is consistent with these
$cMr$ determinations \citep{Merten+15,OS16}. Our $cMr$ is also
consistent with the LOSVD $cMr$ of \citet{GGS16}, which --
  similar to our analysis for this work -- is based on the Jeans
dynamical analysis.  

Our results thus support the $\Lambda$CDM hierarchical model of
cosmological halos formation and evolution at least on the cluster
scales. Our results show that the study of the $cMr$ of clusters can
benefit from the dynamical analysis based on the projected-phase-space
distribution of cluster galaxies, as a complement to other
determinations based on X-ray and lensing techniques and data sets. In
the future, we plan to extend this analysis to higher redshifts and
to the lower end of the mass spectrum of galaxy systems.

\begin{acknowledgements}
  We thank the referee for her/his useful comments that helped
    us improve the quality of our results. We also thank Gary Mamon and
    Barbara Sartoris for useful discussions. \\
  
    We acknowledge financial support from PRIN-INAF 2014.
    B.V. acknowledges the support from an Australian Research
      Council Discovery Early Career Researcher Award (D0028506). \\
    
    This research has made use of data from SDSS DR7.
    Funding for the SDSS and SDSS-II has been provided by the Alfred
    P. Sloan Foundation, the Participating Institutions, the National
    Science Foundation, the U.S. Department of Energy, the National
    Aeronautics and Space Administration, the Japanese Monbukagakusho,
    the Max Planck Society, and the Higher Education Funding Council
    for England. The SDSS website is http://www.sdss.org/.  The SDSS
    is managed by the Astrophysical Research Consortium for the
    Participating Institutions. The Participating Institutions are the
    American Museum of Natural History, Astrophysical Institute
    Potsdam, University of Basel, University of Cambridge, Case
    Western Reserve University, University of Chicago, Drexel
    University, Fermilab, the Institute for Advanced Study, the Japan
    Participation Group, Johns Hopkins University, the Joint Institute
    for Nuclear Astrophysics, the Kavli Institute for Particle
    Astrophysics and Cosmology, the Korean Scientist Group, the
    Chinese Academy of Sciences (LAMOST), Los Alamos National
    Laboratory, the Max-Planck-Institute for Astronomy (MPIA), the
    Max-Planck-Institute for Astrophysics (MPA), New Mexico State
    University, Ohio State University, University of Pittsburgh,
    University of Portsmouth, Princeton University, the United States
    Naval Observatory, and the University of Washington. \\

    This research has made use of data distributed by the NOAO Science
    Archive. NOAO is operated by the Association of Universities for
    Research in Astronomy (AURA), Inc. under a cooperative agreement
    with the National Science Foundation. \\

    This research has made use of the SIMBAD database, operated at
    CDS, Strasbourg, France. \\

    This research has made use of the NASA/IPAC Extragalactic Database
    (NED), which is operated by the Jet Propulsion Laboratory,
    California Institute of Technology, under contract with the
    National Aeronautics and Space Administration.

\end{acknowledgements}

\bibliography{master}

\begin{appendix}

\section{The \texttt{DS+} method of identification of substructures}
\label{a:ds}
The method we developed for identifying cluster members belonging to
substructures is an evolution of the classical method of \citet{DS88},
and we named it \texttt{DS+} after the initials of the authors. 

We start by describing the original test and how it has evolved in time.
The original test looked for the differences $\delta$ of the mean velocity and
velocity dispersion of all possible groups of $\ngr=11$ neighboring
galaxies, from the corresponding cluster global quantities \citep[see
  Eq.~(1) in][]{DS88}.  When the sum of these differences, named
$\Delta$, is much larger than the number of cluster members, $\nm$,
the cluster is likely to contain substructures. The likelihood is
evaluated by Monte Carlo models in which cluster member velocities are
randomly shuffled. 

\citet{Bird94} proposed using $\ngr = \nm^{1/2}$, instead of
11. \citet{Biviano+02} adopted this suggestion and further modified
the original test by also considering the full $\delta$ distribution,
rather than just the sum of the $\delta$s. The authors compared the
observed $\delta$ distribution with Monte Carlo realizations obtained
by azimuthally scrambling the galaxy positions, and identified
statistically significant values of $\delta$, thus pinpointing the
cluster members more likely to reside in substructures.  An additional
modification introduced by \citet{Biviano+02} was to consider only
'cold' groups and to reject as spurious those groups with velocity
dispersions larger than that of the whole cluster.

\citet{Ferrari+03} considered separately the differences in mean
velocity and velocity dispersion, $\delta_v$ and $\delta_{\sigma}$,
respectively. \cite{Girardi+15} evaluated $\delta_{\sigma}$ not with
respect to the cluster global velocity dispersion, but with respect to
the cluster velocity dispersion evaluated at the clustercentric
distance of the group, thus introducing the use of the cluster
velocity dispersion profile in the original method.

Our new \texttt{DS+} method builds upon all these previous
modifications of the original test of \citet{DS88}. Following
\citet{Biviano+02} we adopted the following definitions for $\delta_v,
\delta_{\sigma}$,
\begin{equation}
\delta_v= N_g^{1/2} \, \mid \overline{v_g} \mid \, [(t_{n}-1) \, \slos(R_g)]^{-1},
\end{equation}
and
\begin{equation}
\delta_{\sigma}=[1-\slosg/\slos(R_g)] \, \{1-[(N_g-1)/\chi^+_{N_g-1}]^{1/2}\}^{-1},
\end{equation}
where $N_g$ is the group multiplicity, $R_g$ is the average group distance
from the cluster center, $\overline{v_g}$ is the mean group velocity,
and the Student-$t$ and $\chi^2$ distributions are used to normalize
the differences in units of the uncertainties in the mean velocity and
velocity dispersion \citep[as described in][]{BFG90}, respectively. The mean
velocity of the cluster is null by definition, since we work on
cluster rest-frame velocities.  Following \citet{BFG90} we estimated
the group and cluster velocity dispersions $\slosg$ and
$\slos$ using the biweight estimator for samples of at least 15
galaxies, and the gapper estimator for smaller samples.

The cluster line-of-sight velocity dispersion profile, $\slos(R)$, can
in principle be directly estimated from the cluster member
velocities. However, the result can be very noisy, even for cluster
samples of $\sim 100$ members. We preferred to rely on a theoretical
model. We obtained the cluster $\slos(R)$ by applying the Jeans equation
of dynamical equilibrium and the Abel projection equation
\citep[Eqs.~(8), (9), and (26) in][]{MBB13}, under the assumption of a NFW
mass profile with $\rtwoi$ estimated from the observed total cluster
$\sigv$ (see Sect.~\ref{ss:member}), a concentration given by the
relation of \citet{MDvdB08}, and a velocity anisotropy profile modeled
after the results of numerical simulations \citep{MBM10}. Given that
$\slos$ vary slowly with $R$, the precise choice of the mass profile
concentration has little impact on the results of our analysis.

We considered groups of several possible multiplicities, $\ngr(j)=j
\times 3$, with $j=1, \ldots, k$, where $k$ is the lowest value of
$j$ for which $\ngr(k)>\nm^{1/2}$. In doing this we effectively take
into account that substructures of different richness coexist in a
given cluster. By considering only multiples of triplets we saved in
computing time with little loss of generality.

The sums of each possible group $\delta$s, $\Delta_v \equiv
\sum_{i=1}^{\ngr} \delta_{v}$, and $\Delta_{\sigma} \equiv
\sum_{i=1}^{\ngr} \delta_{\sigma}$, are assigned probabilities via 500
MonteCarlo resamplings. In each of these, we replaced all the cluster
galaxy velocities with random Gaussian draws from a distribution of zero mean
and dispersion equal to $\slos(R_g)$. Groups characterized by
$\Delta_v$ and/or $\Delta_{\sigma}$ probabilities $\leq 0.01$, are
considered significant.

To avoid the case of multiple group assignment for a given galaxy, we
finally eliminated those groups that, even if significant, have one or
more members in common with another group of higher significance
(i.e., lower probability).

By the \texttt{DS+} method we not only identified clusters with
significant presence of substructures, but we identified the
substructures themselves and the galaxies that belong to these substructures. To
assess the accuracy of this new method, extensive tests are needed,
using cluster-size halos extracted from cosmological
simulations. These tests are currently underway and promising
(Zarattini et al., in prep.) but a full analysis of them is beyond the
scope of the present paper. In the future, we plan to
perform a detailed analysis of
the properties of OmegaWINGS cluster substructures and of their
constituent galaxies.

\end{appendix}

\end{document}